\documentclass[aps,pra,twocolumn,nopacs,lettersize,superscriptaddress,nofootinbib]{revtex4-1}
\usepackage{hyperref}
\usepackage{graphicx}
\usepackage{array}
\begin{document}

\title{Configurable microscopic optical potentials for Bose-Einstein condensates using a digital-micromirror device}

\author{G. Gauthier}
\affiliation{ARC Centre of Excellence for Engineered Quantum Systems, University of Queensland, Brisbane, Australia 4072}
\affiliation{School of Mathematics and Physics, University of Queensland, Brisbane, Australia 4072}
\author{I. Lenton}
\affiliation{School of Mathematics and Physics, University of Queensland, Brisbane, Australia 4072}
\author{N. McKay Parry}
\affiliation{ARC Centre of Excellence for Engineered Quantum Systems, University of Queensland, Brisbane, Australia 4072}
\affiliation{School of Mathematics and Physics, University of Queensland, Brisbane, Australia 4072}
\author{M. Baker}
\affiliation{School of Mathematics and Physics, University of Queensland, Brisbane, Australia 4072}
\author{M. J. Davis}
\affiliation{School of Mathematics and Physics, University of Queensland, Brisbane, Australia 4072}
\author{H. Rubinsztein-Dunlop}
\affiliation{ARC Centre of Excellence for Engineered Quantum Systems, University of Queensland, Brisbane, Australia 4072}
\affiliation{School of Mathematics and Physics, University of Queensland, Brisbane, Australia 4072}
\author{T. W. Neely}
\email{t.neely@uq.edu.au}
\affiliation{ARC Centre of Excellence for Engineered Quantum Systems, University of Queensland, Brisbane, Australia 4072}
\affiliation{School of Mathematics and Physics, University of Queensland, Brisbane, Australia 4072}

\date{\today}


\begin{abstract}
Programable spatial light modulators (SLMs) have significantly advanced the configurable optical trapping of particles. Typically, these devices are utilized in the Fourier plane of an optical system, but direct imaging of an amplitude pattern can potentially result in increased simplicity and computational speed. Here we demonstrate high-resolution direct imaging of a digital micromirror device (DMD) at high numerical apertures (NA), which we apply to the optical trapping of a Bose-Einstein condensate (BEC). We utilise a ($1200 \times1920$) pixel DMD and commercially available 0.45 NA microscope objectives, finding that atoms confined in a hybrid optical/magnetic or all-optical potential can be patterned using repulsive blue-detuned (532~nm) light with 630(10)~nm full-width at half-maximum (FWHM) resolution, within 5\% of the diffraction limit. The result is near arbitrary control of the density the BEC without the need for expensive custom optics. We also introduce the technique of time-averaged DMD potentials, demonstrating the ability to produce multiple grayscale levels with minimal heating of the atomic cloud, by utilising the high switching speed (20 kHz maximum) of the DMD. These techniques will enable the realization and control of diverse optical potentials for superfluid dynamics and atomtronics applications with quantum gases. The performance of this system in a direct imaging configuration has wider application for optical trapping at non-trivial NAs.
\end{abstract}
\maketitle

\section{Introduction}
The manipulation of microscopic particles has benefited from the high level of control and measurement provided by optical tweezers. With the technological development of fast configurable spatial light modulators (SLMs) allowing for ever more complex trapping geometries~\cite{grier2003revolution,woerdemann2013advanced,neuman2004optical}, new applications have emerged. For example, sculpted light may have an important role in overcoming multiple light scattering issues in complex biological tissues, and such biomedical applications have only begun to be explored. In particular the development of sculpted light patterns across the image plane, such as the generation of large trapping arrays, could have application to the in vivo trapping of larger objects, such as living cells~\cite{chowdhury2014automated}. 

\begin{figure*}[t!]
\centering\includegraphics[width=.6\textwidth]{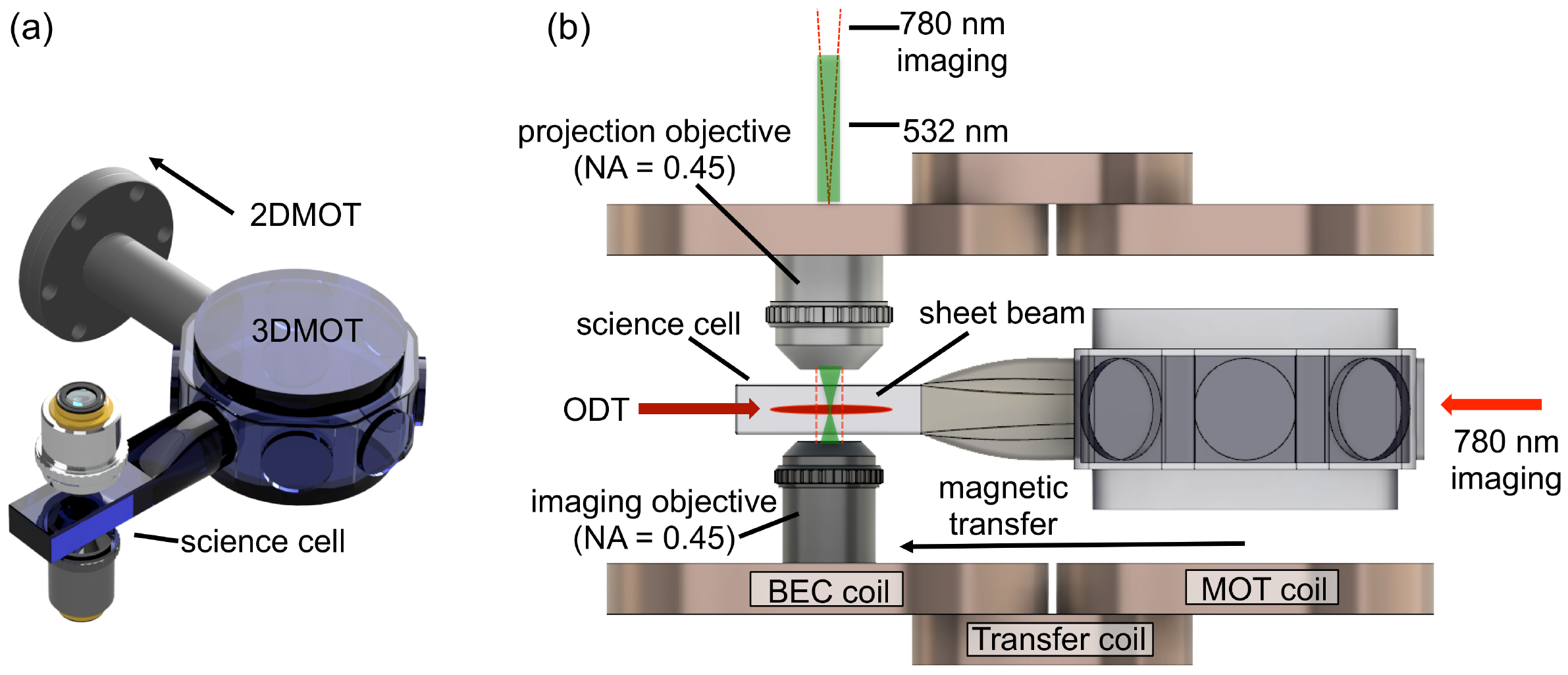}
\caption{Experimental apparatus and BEC production. (a) The 2DMOT (not shown) loads the 3DMOT located in the glass octagon in 5~s. The glass assembly incorporates a commercial Suprasil quartz fluorescence cell (science cell) with 1.25~mm thick walls. (b) After loading into a magnetic trap, the atoms are transferred to the science cell using the coil pairs shown~\cite{parry2014note}. After trapping and evaporation in the optical dipole trap (ODT), the atoms are transferred to a second dipole trap consisting of a red-detuned sheet beam crossed with a vertically-propagating blue-detuned DMD-patterned potential. Time-of-flight (TOF) imaging at $1\times$ magnification is preformed along the horizontal axis, while high magnification in-trap imaging uses the lower microscope system.}
\label{fig_apparatus}
\end{figure*}

In degenerate quantum gases, the push for increasing diversification of optical trapping potentials has led to the adoption of many of the techniques from holographic optical tweezers. SLMs are most often used in the Fourier plane of an optical system, manipulating the phase of an input optical field to produce a configurable pattern in the conjugate trapping plane of the system~\cite{pasienski2008high,gaunt2011robust,harte2014conjugate,nogrette2014single}. These methods have been successfully applied to address and pattern atoms trapped in optical lattices~\cite{preiss2015strongly,fukuhara2013quantum,zupancic2016ultra}, but demonstrations of microscopically configurable trapping potentials have been lacking, with a notable exception being the production of multiple focused spots for the confinement of single atoms~\cite{nogrette2014single}.

An alternative technique to manipulating the phase of the input beam is to instead utilise direct imaging. Though somewhat rarely encountered in optical tweezers, the technique known as \emph{generalised phase constrast} uses the combination of a phase-based SLM and a phase-contrast filter to first create an amplitude pattern in an intermediate image plane, which is then directly reimaged to the optical tweezing plane~\cite{mogensen2000dynamic}. The advantages of this technique are both speed and simplicity -- the desired amplitude pattern can be directly written to the SLM without requiring the calculation of the appropriate hologram. This technique likewise avoids the generation of phase defects and speckle in the imaged pattern that can plague SLMs in the Fourier plane~\cite{pasienski2008high,gaunt2011robust}, while being adaptable to the generation of large numbers of traps~\cite{mogensen2000dynamic,eriksen2002multiple,curtis2002dynamic}. This comes at the cost of the ability to correct wavefront aberrations, but this disadvantage can be mitigated with a well-corrected optical system, as shown here. Another drawback is that the light efficiency is proportional to the fraction of illuminated trap area to maximum trap area.

A more recent addition to the toolkit for producing arbitrary optical potentials has been the digital-micromirror device (DMD). Consisting of (up to) millions of individually addressable mirrors in a compact package, DMDs have the advantage of fast full-frame refresh rates on the order of 20~kHz, $\sim20\times$ that of comparable liquid-crystal based SLMs. DMDs can also be operated in a fixed fashion (DC) as they latch mirrors between reset pulses. Originally developed for digital light processing (DLP), these devices have seen increasing use in laboratory and industrial applications~\cite{dudley2003emerging}. A DMD can be considered a dynamically configurable amplitude mask, which makes it highly suitable for direct imaging applications. DMDs have been used to produce flattop beams for implementation into quantum gas experiments~\cite{liang20091}, incorporated into high-resolution systems for the purpose of single-site addressing in atomic quantum gas microscopes~\cite{preiss2015strongly,fukuhara2013quantum}, used to produce moving lattice potentials~\cite{ha2015roton}, and have recently been utilised to produce target-shaped traps~\cite{kumar2016minimally}. DMDs may be used in either the Fourier plane~\cite{preiss2015strongly,zupancic2016ultra,stilgoe2016interpretation,mirhosseini2013rapid,liu2015generalized} or directly imaged~\cite{fukuhara2013quantum,kumar2016minimally}. 

\begin{figure*}[t!]
\vspace{12pt}
\centering\includegraphics[width=.65\textwidth]{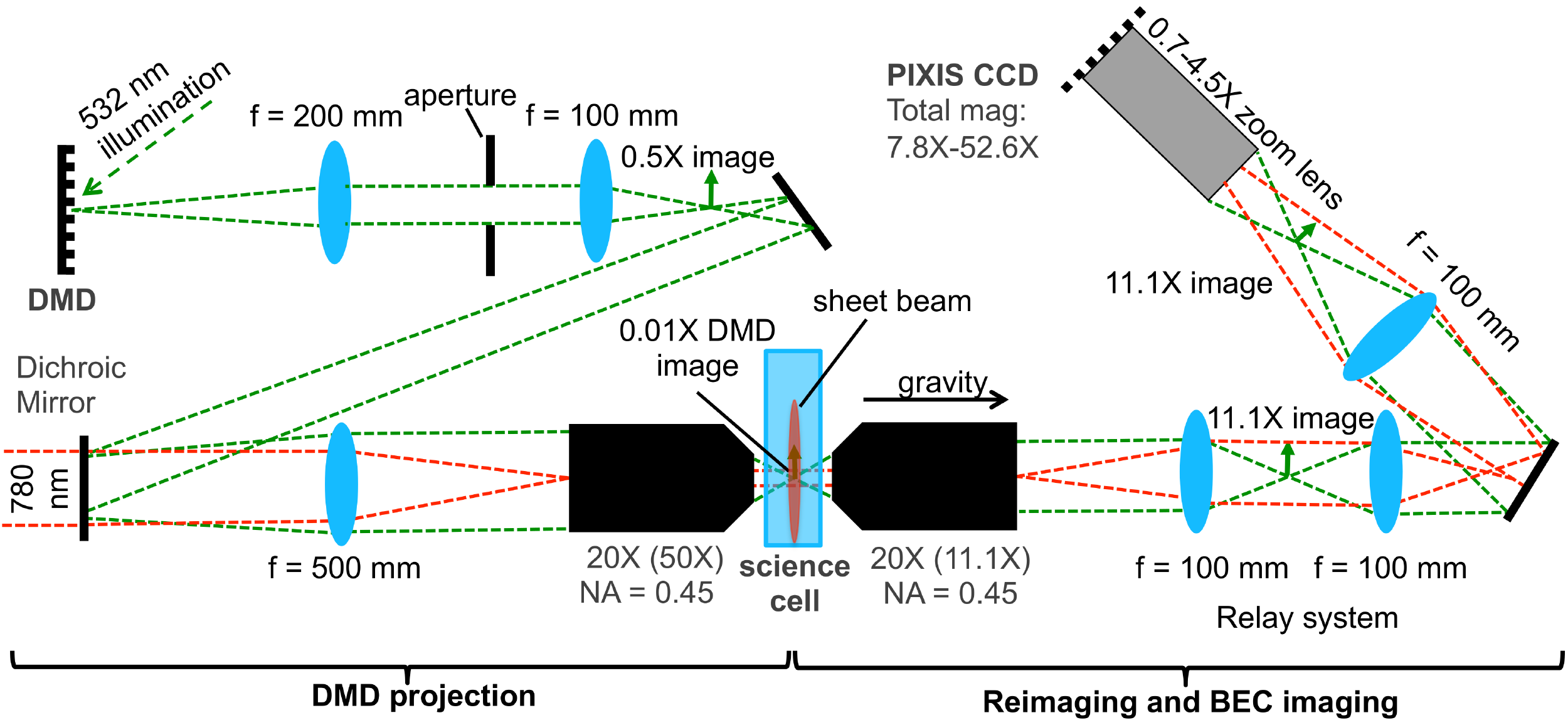}
\caption{DMD projection system and reimaging/BEC imaging system. The DMD is imaged on to the atom plane with $0.01\times$ magnification ($100\times$ minification), and combined with a red-detuned TEM$_{00}$ sheet. The bottom system similarly produces an image of the BEC or DMD pattern, with adjustable magnification from $7.77\times$ -- $52.6\times$. The lens relay system will be used for a future implementation of darkground imaging~\cite{wilson2015situ,pappa2011ultra}.}
\label{fig_imaging}
\end{figure*}

We demonstrate here the utility of direct imaging of a DMD at high numerical apertures (NA) for optical trapping, which we apply to trapping of a BEC. Our optical system has the major advantage of using commercially available optics and microscope objectives external to our glass vacuum chamber, and corrected for the relatively thin 1.25~mm thick walls. With this system we demonstrate patterning of potentials with an upper-bound resolution of 630(10)~nm FWHM at 532~nm illumination, within 5\% of the diffraction limit for our 0.45 ~NA objective and $\sim5.5\times$ improved on recently reported results~\cite{kumar2016minimally,endnote1} which utilised a relatively low $\sim0.08$~NA imaging system. The high-resolution potentials appear robust to tilts and misalignments of the objective and glass walls, in contrast to other cold atom experiments~\cite{zimmermann2011high}. These patterns have an image extent of $130~\mu\textrm{m}\times207\mu\textrm{m}$ in the atom plane, which allows nearly arbitrary sculpting of the optical potential and corresponding BEC. Subsequent imaging at the 780~nm resonant wavelength of $^{87}$Rb achieves a submicron resolution of 960(80)~nm FWHM, within 8\% of the diffraction limit. Our DMD allows the storage of 13,889 frames and has a full-frame frame rate ranging from DC to 20~kHz, enabling diverse and dynamically configurable potentials. We utilise this rapid switching rate to introduce the use of DMDs for producing time-averaged potentials,  which have been previously produced from rapidly scanning beams~\cite{schnelle2008versatile,henderson2009experimental,bell2016}. We find that modulation frequencies above $\sim3$~kHz produce negligible heating of the atoms while allowing the production of six grayscale levels. This technique can be combined with binary error-diffusion (halftoning)~\cite{liang20091} to increase the number of grayscale levels available.

\section{Apparatus and BEC production}
Our apparatus consists of a three-part vacuum system, comprising a two-dimensional magneto-optical trap (2DMOT), three-dimensional MOT (3DMOT) and attached science cell cuvette (see Fig.~\ref{fig_apparatus}). The cell is made of Suprasil quartz and has 1.25 mm thick walls with an external broadband anti-reflection coating. The 2DMOT is separated from the 3DMOT and science cell sections with a 12~mm long differential pumping tube with a 1.2~mm inner diameter. This results in a pressure differential of $\sim15,000$, with a pressure of $\sim 5\times10^{-12}$~Torr and $10^{-7}$~Torr on the 3DMOT and 2DMOT sides respectively. The 2DMOT loads $3\times 10^9$ $^{87}$Rb atoms into the 3DMOT in 5~s. The subsequent BEC production closely follows the methodology of Ref.~\cite{lin2009rapid}. After a 40~ms compressed MOT stage (CMOT)~\cite{petrich1994behavior}, the atoms are optically pumped to the $F = 1$ manifold and captured in a 100~G/cm quadrupole magnetic field. Using two additional anti-Helmoltz coil pairs~\cite{parry2014note} (see Fig.~\ref{fig_apparatus}), the atoms are adiabatically transferred to the science cell in $\sim0.8$~s~\cite{greiner2001magnetic}. The quadrupole field is subsequently increased to 145~G/cm, and evaporation occurs over 4~s by driving the microwave transition from $|F=1,m_F = -1\rangle \rightarrow |F=2,m_F = -1\rangle$, resulting in a cloud of $1.4\times10^8$ atoms at a temperature of $20~\mu$K. The field is then relaxed to 27 G/cm, loading $4.5\times10^7$ atoms at $4.5~\mu$K into a red-detuned 1064~nm beam with a $95~\mu$m waist, aligned $\sim80~\mu$m below the quadrupole zero, and with a potential depth of 110~$\mu\textrm{K}\times\textrm{k}_{\textrm{B}}$. We find that the cancellation of stray magnetic fields is important for efficient loading of the dipole trap. Optical evaporation is performed by reducing the beam intensity, resulting in a temperature of $\sim450$~nK, with a critical temperature for condensation of $300$~nK. If the optical evaporation is continued, BECs of $5\times10^6$ atoms can be produced in this trap. Instead, we load the cold thermal atoms into a second 1064 nm red-detuned beam focused into a sheet with radial and vertical waists ($w_r,w_z$) = (1~mm, 8.5~$\mu$m) and a trap depth of 6.4~$\mu\textrm{K}\times\textrm{k}_{\textrm{B}}$. Subsequent evaporation in the sheet-magnetic hybrid trap results in a BEC of $\sim4\times10^6$ atoms in a highly flattened geometry, with final trapping frequencies of $(\omega_r,\omega_z) = 2\pi \times(20, 310)$~Hz. After evaporating in the sheet hybrid trap, we subsequently remove the magnetic field, while ramping up the sheet potential, resulting in all-optical confinement with $(\omega_r,\omega_z) = 2 \pi \times (6, 380)$~Hz. The addition of the DMD potential to the sheet (see below) can dominate the weak radial harmonic confinement of this trap, producing nearly uniform atomic distributions.

\section{DMD projection}
In order to produce arbitrarily patterned BECs, we directly image a DMD (Visitech LUXBEAM 4600 WUXGA 1200$\times$1920 pixels) to the atom plane by illuminating the surface with spatially-filtered 532~nm light as shown in Fig.~\ref{fig_imaging}. This results in approximately Gaussian illumination with $w_0 = 12.3$~mm and maximum DMD diffraction efficiency of $\sim$~40\%. An intermediate image of the DMD with $0.5\times$ magnification is initially produced using a matched achromat imaging system. An aperture at the Fourier plane of this system removes spurious DMD diffraction orders, which otherwise contribute stray light to the dark regions of the imaged patterns and reduce the trap lifetime. Subsequently, the DMD is imaged to the atom plane using a 500~mm achromat and Nikon CFI L PLAN EPI 20XCR infinity corrected objective, resulting in $0.02\times$ magnification. The total system thus images the DMD to the atom plane with a final minimization factor of 100 ($0.01\times$ magnification). The objective has a numerical aperture (NA) of 0.45 and an adjustable glass correction collar. We have found the correction collar to be important in obtaining high-resolution patterns inside the science cell, but the system is relatively insensitive to the precise angle of the objective and only requires coarse tuning.

\begin{figure}[t!]
\centering\includegraphics[width=.8\columnwidth]{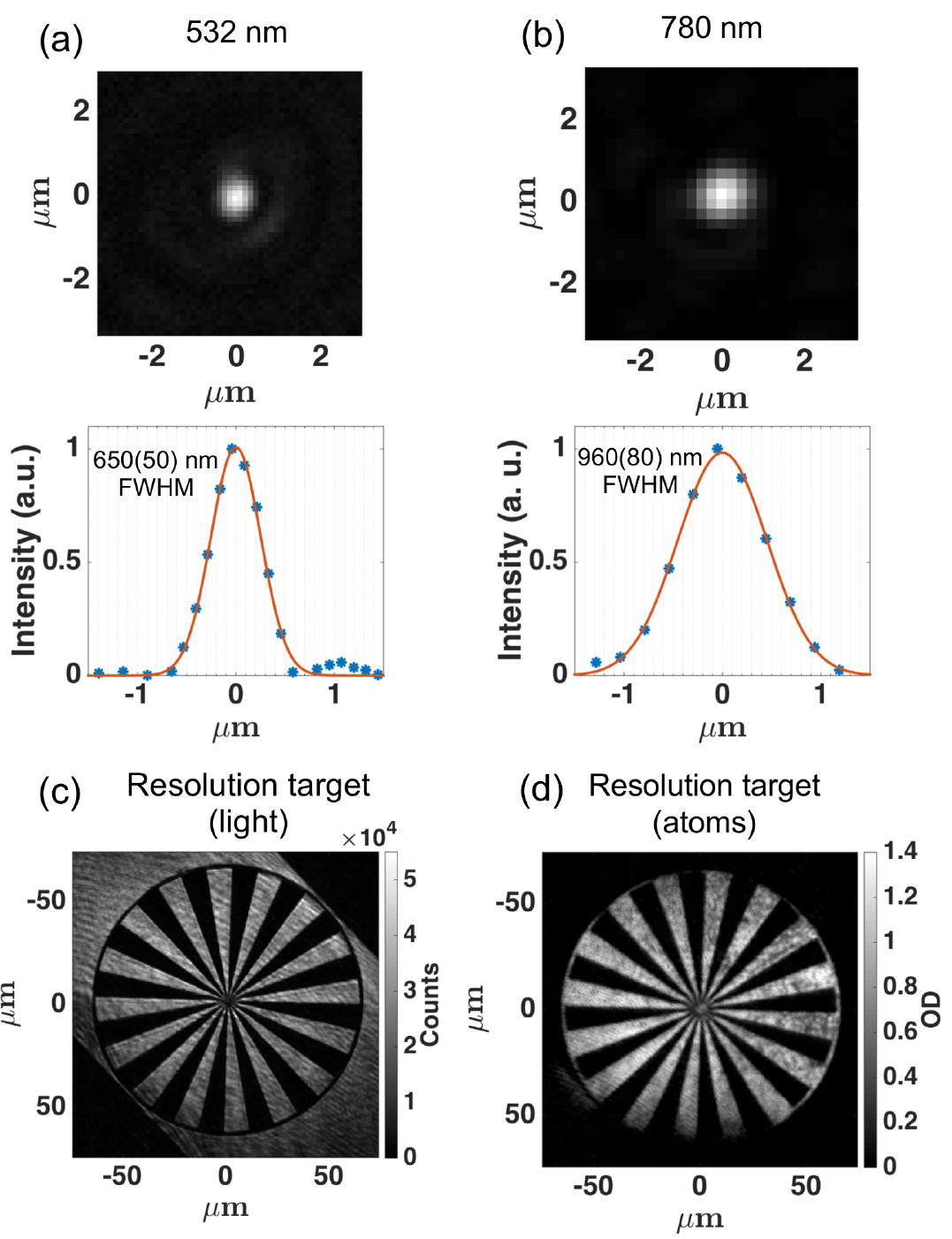}
\caption{(a) A single DMD mirror imaged on to the camera with $105\times$ magnification and 532~nm illumination, resulting in 650(50)~nm FWHM ($\omega_0 \sim 550$~nm). The FWHM was determined through 100 1D fits of the image at varying angles through 180 degrees; a single 1D fit is shown here for illustrative purposes. (b) A single DMD mirror imaged with $52.6\times$ magnification and 780~nm illumination, the imaging wavelength for $^{87}$Rb, resulting in 960(80)~nm FWHM ($\omega_0 \sim 814$~nm). (c) Siemens star resolution target, as projected and reimaged with 532~nm light. (d) Siemens star imprinted on to the atomic density and averaged over 10 runs of the experiment. Quantitative analysis of the resolution targets is presented in Fig.~\ref{fig_MTF}.}
\label{psf_fig}
\end{figure}

As the DMD mirrors have only two states --- on and off --- the resulting image is binary. However, given the limited spatial resolution of a typical optical system, the technique know as error-diffusion or halftoning can be used to produce intensity gradients~\cite{liang20091,kumar2016minimally,floyd1976adaptive}. We determine the resolution of our system by first turning on a single mirror of the DMD and projecting it on to the atom plane. The mirror pitch is 10.8~$\mu$m, so the minification factor of 100 results in an imaged mirror width of 108~nm, below the theoretical resolution limit of the top microscope system (605~nm FWHM at 532~nm illumination). As the single mirror is not resolvable, its image at the atom plane approximates the point-spread function (PSF) of the DMD imaging system. We reimage this focused spot on the camera with the bottom microscope system. Accounting for the magnification factor of the reimaging system results in a 650(50)~nm FWHM peak in the atom plane, as shown in Fig.~\ref{psf_fig}(a). This value is an upper limit, as it convolves any abberations of the reimaging system into the estimate of the spot size at the atom plane. Back-propagating this resolution element to the DMD location gives a 65~$\mu$m spot which spans a $\sim6\times6$ block of mirrors; multiple mirrors will thus contribute to the resolution spot in the atom plane~\cite{liang20091}. We therefore can use error-diffusion to control the light intensity at the atom plane, as shown in Fig.~\ref{fig_dmd_patterns}(b).

\section{Atom trapping and imaging}
The right half of Fig.~\ref{fig_imaging} illustrates the bottom microscope imaging system. A second microscope objective (Olympus LCPLN20XIR) similar to the projection objective with a 0.45 NA and correction collar, but with near-infrared coatings, forms the first part of the imaging system. Collimated light at 780~nm is produced through the top objective, resulting in an imaging beam 1.1~mm in diameter. We then incorporate a lens relay system to allow us to implement dark-field imaging in the future~\cite{wilson2015situ,pappa2011ultra}. Finally, an intermediate image is formed with $11.1\times$ magnification which is then reimaged to the CCD camera (PIXIS 1024B) with the adjustable VZM450 zoom lens ($0.70\times$ -- $4.74\times$). We therefore produce absorption images of the condensate in the final trap configurations with a total zoom of ($7.77\times$ -- $52.6\times$). By again turning on a single mirror of the DMD, but illuminating it at 780~nm, we can estimate the PSF at this wavelength, as shown in Fig.~\ref{psf_fig}(b). This results in a 960(80)~nm FWHM peak, consistent with the measurement at 532~nm after accounting for the increased wavelength.

\begin{figure*}[p!]
\centering\includegraphics[width=.9\textwidth]{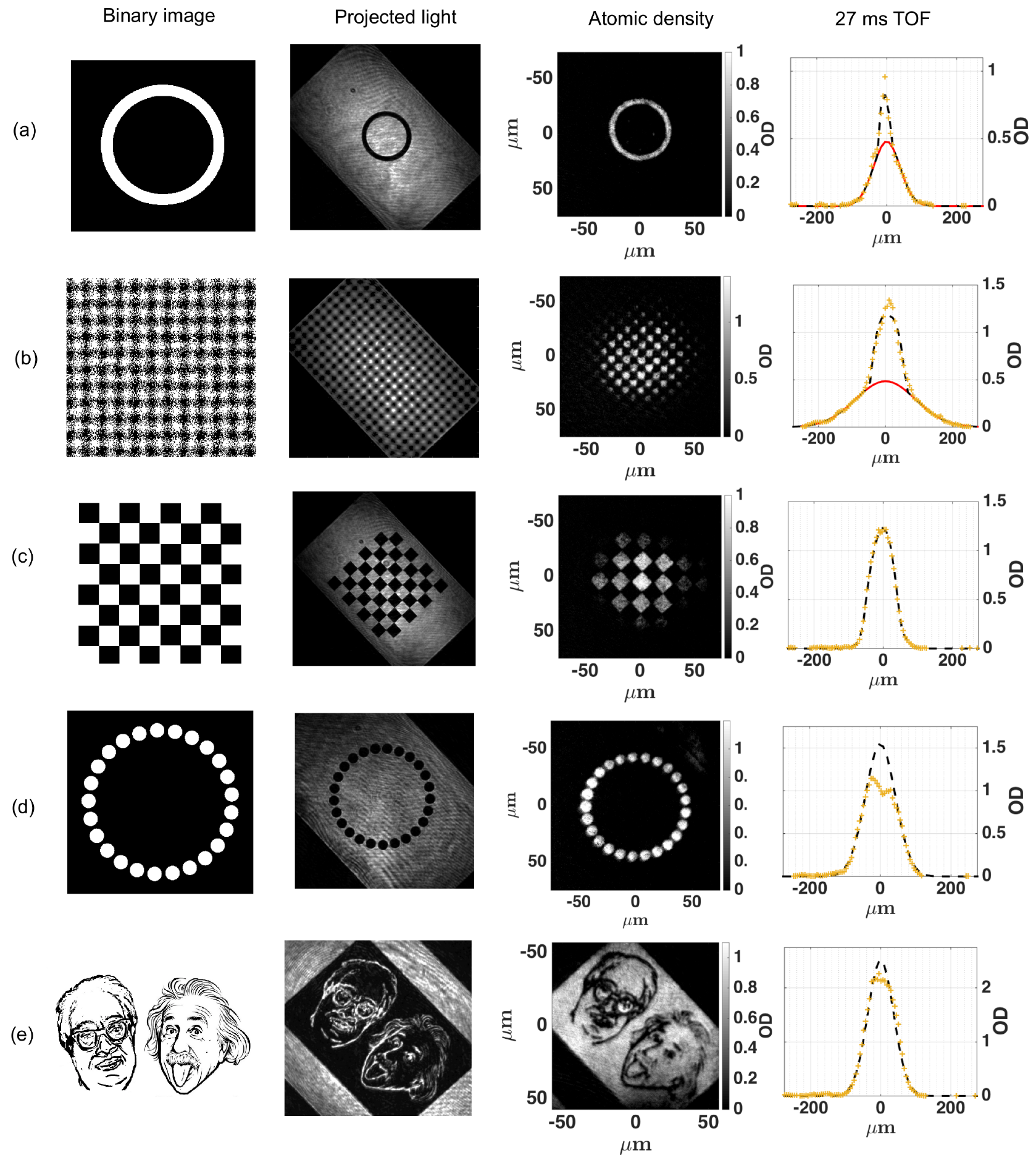}
\caption{DMD patterned optical traps and resulting resonant absorption images of atom distributions. Atoms are repelled from bright regions of the projected pattern, which is the inverse of the binary image applied to the DMD. We image the atoms immediately after turning off the optical trapping potentials, with a magnification of $52.6\times$. Bright areas represent regions of high atomic density and optical depth (OD). (a) Ring trap potential from a single experimental realization. Time of flight analysis (TOF) gives $N =1.3\times10^5$  atoms with matter-wave interference leading to the appearance of a central peak~\cite{mathey2010phase}. (b) Lattice pattern produced by applying a Floyd-Steinberg error diffusion algorithm~\cite{floyd1976adaptive} to a 8-bit image of a sinusoidal lattice with $10~\mu$m period, shown with a single realization of a BEC; $N = 3.7\times 10^5$ and the BEC fraction is 34\%. This image was produced by leaving the magnetic trap on, resulting in $\omega_r = 2\pi\times$20~Hz harmonic radial confinement. (c) Checkerboard pattern applied to the atoms, imaged with a single shot. Additional evaporation after transfer to the all-optical trap results in a nearly pure BEC in TOF with $N = 2.4\times10^5$ atoms. (d) Ring lattice of 25 sites, with ring radius of 43.2~$\mu$m, and site radius of 4.32~$\mu$m, with $N = 3.16\times10^5$ atoms. ODs above $\sim2.5$ are below the signal to noise threshold of the horizontal-imaging, leading to slight undercounting of the atoms in this case. (e) Artistic impressions of Bose and Einstein~\cite{einstein} applied to a nearly pure Bose-Einstein condensate of $N = 5.2\times10^5$ atoms, averaged over 5 experimental runs.}
\label{fig_dmd_patterns}
\end{figure*}

The resulting high resolution DMD patterns produce a repulsive potential for the atoms trapped in the sheet-magnetic hybrid. We ramp on the intensity of the laser illuminating the DMD near the end of the final evaporation sequence in the sheet-hybrid trap to a value $\sim1.2\mu$, where $\mu$ is the chemical potential of the BEC in the final trapping potential. In the case of a 80~$\mu$m~$\times$~50~$\mu$m rectangular trapping geometry, the chemical potential calculated using the measured trap frequencies and 700,000 atoms is $\sim50~\textrm{nK}\times\textrm{k}_\textrm{B}$. Subsequently, the sheet power is linearly increased over 500~ms to keep the atoms trapped vertically while the magnetic field is linearly ramped to zero, increasing the vertical trapping frequency to $\omega \sim2\pi \times380$~Hz. The light power incident on the DMD can also be increased in this step. However, we have observed a decrease in the BEC lifetime with increasing intensity,  with measured time constants of (18.2, 10.9, 9.1, 8.2)~s for a 80~$\mu$m~$\times$~50~$\mu$m rectangular DMD box potential with peaks of (1.2$\mu$, 2.4$\mu$, 4.8$\mu$, 7.2$\mu$) respectively. Similar intensity-dependent lifetimes were observed for different trap geometries. As the atoms are confined to areas of low DMD light intensity, we believe residual vibrational coupling in the DMD optical path results in parametric heating of the cloud.

Several high-resolution atom-trap configurations are shown in Fig.~\ref{fig_dmd_patterns}, emphasizing the wide range of possible patterns. These are imaged immediately after turning off the optical trap. Alternatively, by imaging from the side after 27~ms time of flight (TOF) expansion, we can determine atom number and BEC fraction for the different distributions. We note that we have used the same evaporation ramp for each of these configurations --- we expect that a near pure BEC can be achieved for each with optimization. We generally observe lower condensate fractions for smaller enclosed areas for similar optical evaporation profiles.

We briefly describe examples of the trapping potentials we have generated. Figure~\ref{fig_dmd_patterns}(a) demonstrates a ring trap, of interest for atom interferometry~\cite{gupta2005bose} and studies of phase slips and persistent currents~\cite{eckel2014interferometric,corman2014quench}. Figure~\ref{fig_dmd_patterns}(b) shows a 10~$\mu$m-period optical lattice. The binary DMD pattern was generated by applying a Floyd-Steinberg error diffusion algorithm~\cite{floyd1976adaptive} to an 8-bit grayscale image of a sinusoidal lattice. As we have not included a boundary to the lattice in the DMD pattern, we leave the magnetic field on and retain a $2\pi\times20$~Hz radial trapping frequency, producing a more symmetrically filled lattice. Figure~\ref{fig_dmd_patterns}(c) demonstrates a checkerboard pattern applied to the BEC, and the evaporation ramp in this case results in a BEC with negligible thermal component. This pattern emphasizes the sharp features in the atomic distribution resulting from the highly resolved features of the DMD pattern. Ring lattices have generated wide interest~\cite{amico2005quantum}, and Fig.~\ref{fig_dmd_patterns}(d) demonstrates a ring lattice of 25 sites. Finally, in Fig.~\ref{fig_dmd_patterns}(e) we project artistic impressions of Bose and Einstein into a nearly pure BEC, resulting in a "Bose-Einstein" Bose-Einstein condensate  and demonstrating our ability to create arbitrary potentials. This image is an average over 5 experimental runs (30~s experimental cycle) and is similar to Fig.~\ref{psf_fig}(d) which uses a 10-run average. These images demonstrate the repeatability of the high-resolution patterned BECs. The integrated OD TOF cross sections in Fig.~\ref{fig_dmd_patterns}(d-e) deviate from Gaussian fit at their center. In the case of Fig.~\ref{fig_dmd_patterns}(d), this is due to uneven density distribution of atoms across the ring lattice and insufficient spatial overlap, or incoherent overlap of the lattice sites after only 27~ms TOF. In the case of Fig.~\ref{fig_dmd_patterns}(e), this deviation comes from to the high atomic densities leading to ODs above $\sim2.5$ which  are below the signal to noise threshold of the horizontal-imaging. In each of these cases, the limited expansion of the cloud, when compared with the thermal component of Fig. 4(b), along with the lack of a bimodal distribution, indicate a relatively pure BEC state.

\section{Measuring the modulation transfer function from light and atomic distributions}
\begin{figure}[t!]
\centering\includegraphics[width=.8\columnwidth]{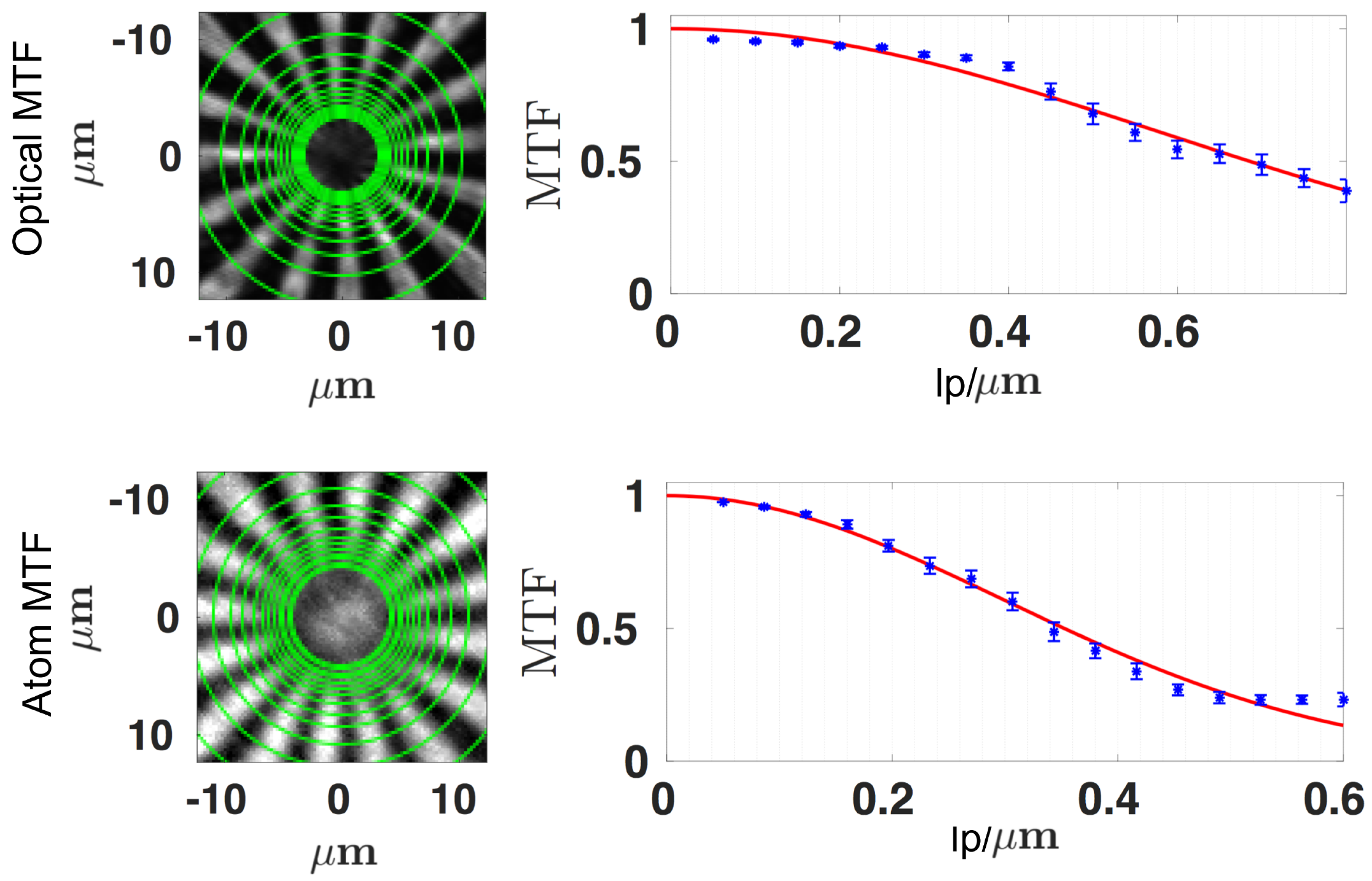}
\caption{Modulation transfer function (MTF) analysis. Left: zoomed versions of the Siemens star images in Fig.~\ref{psf_fig}, with the top row representing the optical pattern at 532~nm illumination, and the bottom an average atomic density in the pattern, imaged with resonant 780~nm light. The green circles indicate radii of equally seperated spatial frequencies used to generate the corresponding MTF plots by calculating contrast along the circular path. Right: the FWHM of the optical pattern is $1.39(0.02)$~lp/$\mu$m, corresponding to a PSF FWHM of 630(10)~nm. The atomic density MTF FWHM is $0.71(0.02)$~lp/$\mu$m, corresponding to a PSF FWHM of 1250(20)~nm.}
\label{fig_MTF}
\end{figure}

For a more quantitative characterisation of the optical systems~\cite{horstmeyer2016standardizing}, we produced a binary Siemens star resolution target on the DMD, imaged the projected light pattern, and imaged BECs loaded in to the pattern, as shown in Fig.~\ref{psf_fig}(c,d). These patterns can then be analyzed to determine the modulation transfer function (MTF) of the total optical system. This is accomplished using the protocol illustrated in Fig.~\ref{fig_MTF}. We first determine circular paths around the Siemens star, with frequency spacings $\Delta = (0.047, 0.034)$ line-pairs (lp)/$\mu$m, for the (light, atom) Siemens star. A contrast value is found for each of the 16 adjacent bright and dark spoke pairs, and the average contrast is calculated for each circular path. As the MTF is the Fourier transform of the PSF, which we estimate with a Gaussian fit, we extract a FWHM from a Gaussian fit of the measured MTF. These quantities are then related by $\textrm{FWHM}_{\textrm{PSF}} = 4\textrm{ln}(2)(\pi\textrm{FWHM}_{\textrm{MTF}})^{-1}$, allowing us to compare the PSF FWHM extracted from the MTF with that of the single mirror image. For the Siemens star with 532~nm light, we find a MTF FWHM of $1.39(0.02)$~lp/$\mu$m, corresponding to a PSF FWHM of 630(10)~nm, in agreement with the 650(50)~nm measurement of the single mirror image. When imaging a BEC in the Siemens star pattern, we find a MTF FWHM of $0.71(0.02)$~lp/$\mu$m, corresponding to a PSF FWHM of 1250(20)~nm, larger than the 960(80)~nm single mirror image at 780~nm illumination.  We believe that this $\sim30\%$ increase is consistent with atom diffusion due to photon recoil during the 10~$\mu$s repump and resonant imaging pulse~\cite{muessel2013optimized}.

\section{Time-averaged potentials with DMDs}
The DMD is capable of switching mirrors from DC to the specified maximum frequency of 20~kHz, which we have verified through photodiode measurement. This wide modulation range, along with the ability to store 13,889 frames on the device, enables a high level of dynamic control. At low modulation frequencies this allows adiabatic deformation of the DMD potentials. At the other extreme, high frequency switching can be utilised for quickly quenching the potential geometry. Furthermore, the painted-potentials technique is possible, where an average dipole force is produced through rapid modulation of the optical field~\cite{schnelle2008versatile,henderson2009experimental}. This suggests pulse-width-modulation (PWM) as an alternative technique to error diffusion for producing grayscale levels, analogous to the techniques used in DLP.

\begin{figure}[t!]
\centering\includegraphics[width=.9\columnwidth]{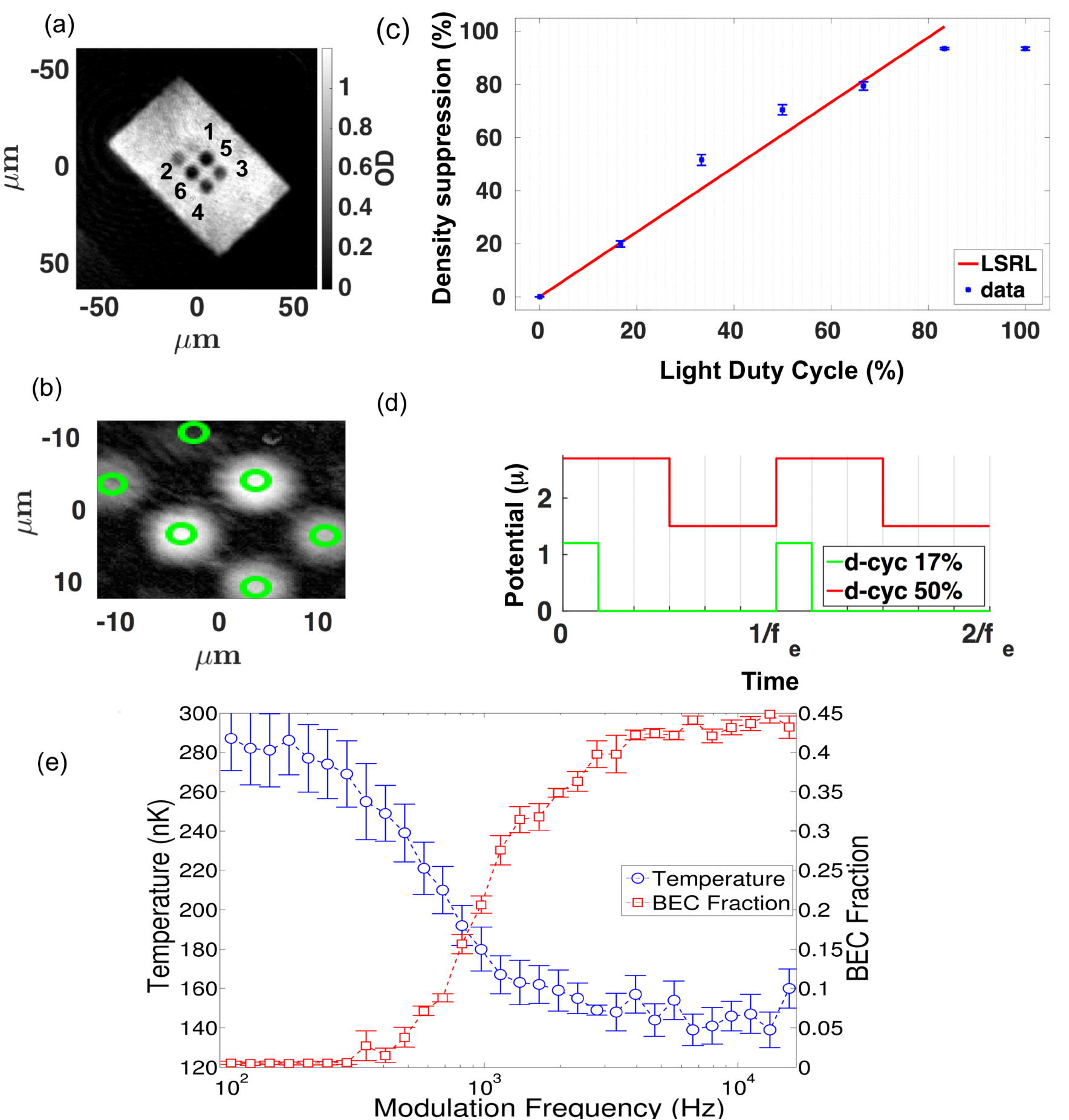}
\caption{(a) Grayscale time-averaged pattern applied to the atoms, averaged over 10 sequential images. Each of the numbered patterns correspond to the fractional duty cycle of 2.75~kHz, with 6 being the maximum (see text), corresponding to potential depth of $1.2~\mu$. (b) Images used for analysis of the gray levels. The grayscale image was subtracted from the average atomic background with no barriers present. The mean density and standard deviation were calculated over the circles as indicated. (c) Gray levels achieved through this process. A least squares regression line (LSRL) is indicated. As the optical potential of point 6 exceeds the condensate chemical potential it was excluded from the fit. Image background and untrapped atoms result in an apparent maximum density suppression of 96\%. (d) Two example pulse width modulations, corresponding to patterns 1, the minimum pulse, and 3, a 50\% duty cycle, offset vertically for clarity. The envelope frequency $f_e = 2.75$~kHz and carrier frequency $f_c = 16.5$~kHz divisions are indicated. (e) Turning on and off barrier 3 only, with varying frequency, for a total modulation time of 500~cycles provides an estimate of the heating rate from mirror switching. Rates above $\sim3$~kHz appear to have a negligible heating effect over this modulation period.}
\label{fig_time_avg}
\end{figure}

To explore the suitability of this system for producing time averaged potentials, we produced the DMD pattern shown in Fig.~\ref{fig_time_avg}, consisting of an array of six 8~$\mu$m diameter barriers contained in a 50~$\mu$m $\times$ 80~$\mu$m rectangle. PWM, with the DMD running at an envelope frequency of $f_e = 2.75$~kHz, was utilised, and the leading edge of each pulse was fixed. The hexagonal array within the rectangle was modulated with varying duty cycle over six levels by subdividing the carrier pulse into six modulation divisions, with the maximum duty cycle corresponding to all divisions turned on, and the minimum corresponding to one division turned on. The carrier frequency thus used was $f_c = 16.5$~kHz. We illuminated the DMD with light corresponding to a trap depth of 1.2$\mu$.  The corresponding relative atomic densities are shown in Fig.~\ref{fig_time_avg}(c).

To investigate the limits of this technique, we measured the effect on temperature and BEC fraction due to modulation of a single barrier (labeled "3" in Fig.~\ref{fig_time_avg}) which had a 50\% duty cycle. After first forming the BEC in the rectangle in the absence of any internal barrier, we performed a modulation for 500 cycles at fixed frequencies ranging from 100~Hz to 16~kHz. We observed a decreasing effect as the frequencies are increased due to net energy gain per cycle decreasing since the barrier oscillates on a smaller time scale than atom diffusion timescales; above $\sim3$~kHz the effect is negligible.

\section{Conclusions}
DMD devices combined with commercially-available glass-corrected objectives present a powerful technique for both the microscopic patterning of quantum gases and more general optical trapping. By utilising a commercial fluorescence cell with relatively thin 1.25~mm walls, expensive custom-built objectives can be avoided while still achieving high resolution. We are able to estimate a 630(10)~nm FWHM PSF for our DMD projection system at 532 nm illumination, in agreement with the single-mirror analysis. A similar estimate of the PSF at 780~nm using a single mirror gives 960(80)~nm FWHM, while MTF analysis of an atomic resolution target gives 1250(20)~nm FWHM. We believe the broadening is due to diffusion of the atoms during the 10~$\mu$s repump and resonant imaging pulses. We note that off-resonant imaging techniques such as darkground or Faraday imaging~\cite{wilson2015situ,pappa2011ultra} might allow atomic distributions to be imaged with the full resolution. 

Commercial DMD devices are now well-suited for implementation in both quantum gas experiments and other optical trapping applications. The painted-potentials technique, in combination with error diffusion methods, can generate grayscale potentials with more than one hundred levels. The ability to easily store and trigger 13,889 frames at high speed shows great promise for the dynamic control of these microscopic potentials. Furthermore, the ability to nearly instantaneously quench the potential geometry has many potential applications, such as investigations of superfluid transport between reservoirs~\cite{lee2013analogs,lee2015contact}, and a demonstration of the superfluid fountain effect in BECs~\cite{karpiuk2012superfluid}. These techniques may prove useful in the production of superfluid turbulence in BECs, where, for example, the production and control of vortex pairs using moving potential barriers has recently been demonstrated~\cite{samson2015deterministic,tw2012experimental}. Configurable trap geometries will be necessary for designing atomtronic circuits~\cite{seaman2007atomtronics}. As the DMD is largely wavelength insensitive (except in terms of diffraction efficiency), the possibilities for use at multiple wavelengths are intriguing. This capability could be utilised for the production of species dependent potentials~\cite{catani2009entropy}. Concurrent work by the Munich group has shown the usefulness of high resolution direct imaging (600~nm FWHM) of a DMD in creation of a two-dimensional disordered lattice for the exploration many body localization transitions~\cite{Choi2016exploring}.

More generally, our results demonstrate the utility of direct imaging of an DMD. With a well-corrected optical system, performance close to the diffraction limit can still be achieved at a non-trivial 0.45~NA. The computational simplicity of direct imaging may prove an advantage in applications that require reconfiguration in real-time; the high operating speed of the DMD furthermore facilitates this. These results are also applicable to optical trapping beyond quantum gases and in particular may be advantageous for the production of numerous traps for confining arrays of particles~\cite{chowdhury2014automated,eriksen2002multiple,curtis2002dynamic}.

\section*{Funding Information}
Australian Research Council (ARC) (CE1101013, DP160102085).

\section*{Acknowledgments} 
We would like to acknowledge the contributions of Thomas Carey to the initial technical development of the BEC apparatus, and Alexander Stilgoe for useful discussions.


\end{document}